\documentstyle[11pt,paspconf]{article}

\def\sb{\ifmmode{\;{\rm mag}\;{\rm arcsec}^{-2}}\else{~mag~arcsec$^{-2}$}\fi}
\def\linsb{\ifmmode{\;L_{\sun}{\rm pc}^{-2}}\else{~$L_{\sun}{\rm pc}^{-2}$}\fi}
\def\csb{\ifmmode{\mu_0}\else{$\mu_0$}\fi}
\def\lincsb{\ifmmode{\Sigma_0}\else{$\Sigma_0$}\fi}
\def\solar{\ifmmode{_{\sun}}\else{$_{\sun}$}\fi}

\def\etal{{et al}.\ }
\def\MLo{\ifmmode{\Upsilon}\else{$\Upsilon$}\fi}
\def\ML*{\ifmmode{\Upsilon_*}\else{$\Upsilon_*$}\fi}
\def\MLT{\ifmmode{\Upsilon_T}\else{$\Upsilon_T$}\fi}
\def\mass{\ifmmode{\cal M}\else{${\cal M}$}\fi}
\def\kms{\ifmmode{\;{\rm km}{\rm s}^{-1}}\else{~${\rm km}{\rm s}^{-1}$}\fi}

\begin{document}

\title{The Baryon Fraction Distribution \\ and the Tully-Fisher Relation}
\author{Stacy McGaugh}
\affil{Department of Terrestrial Magnetism, Carnegie Institution of
Washington, 5241 Broad Branch Road, NW, Washington, DC 20015}
\author{Erwin de Blok}
\affil{Kapteyn Astronomical Institute,
Postbus 800, 9700 AV Groningen, The Netherlands}

\begin{abstract}
A number of observations strongly suggest that the baryon fraction is
not a universal constant.  One obvious interpretation
is that there is some distribution of $f_b$, and the different observations
sample different portions of the distribution.  However, the small
intrinsic scatter in the Tully-Fisher relation requires that the
baryon fraction be very nearly universal.  It is not easy to resolve this
paradox in the framework of the standard picture.
\end{abstract}

\keywords{Dark Matter,Tully-Fisher Relation}

\section{The Baryon Fraction Must Vary}

There now exist a number of observations which indicate that the
ratio of luminous to dark mass is not the same for all systems.
Rather than the single universal baryon fraction that we have nominally assumed,
$f_b$ seems to have a broad distribution.  Some of the observations
suggesting this situation include satellite galaxies, tidal tails,
X-ray clusters of galaxies, and low surface brightness galaxies.

The satellite studies
of Zaritsky \etal (1994, 1997) imply that the halos of $L^*$ galaxies
are very large and massive.  On the other hand, the morphology of tidal
tails is very difficult to reproduce unless the mass of halos does not
exceed the disk mass by more than a factor of 10 (Dubinski \etal 1996).
This result is inconsistent with the satellites result by a factor of
$\sim 2$.  One might be tempted to equivocate at this level, but
the result should not be lightly dismissed (Mihos \etal 1997).

The baryon fraction is directly estimated in X-ray clusters
(White \etal 1993; Evrard \etal 1996).  These indicate $f_b \sim 0.1$
with significant (factor of two) scatter which is argued to be real
(White \& Fabian 1995).  In contrast, a stringent limit is placed by
the rotation curves of the most dark matter dominated galaxies:
$f_b < 0.05$ (de Blok \& McGaugh 1997).  This is not really consistent
with the cluster result, and attempting to fit the rotation curves with
NFW halos (Navarro \etal 1996) requires even lower baryon fractions of
0.01 -- 0.02.  This differs from clusters by nearly an order of magnitude,
and is very difficult to explain away.  Mass expulsion of baryons is often
invoked, but this scenario predicts that the gas should be swept away.
The actual galaxies are in fact quite gas rich.  Other
explanations could be offered, but most are rather hand-waving and
lack predictive power.

Another approach is to suppose that there is a
distribution of baryon fractions.  The apparently contradictory observations
might then be reconciled:  they simply happen to sample different portions
of the distribution.  However, this apparently reasonable approach has a
serious problem in explaining the Tully-Fisher relation.

\section{The Baryon Fraction Must Not Vary}

The traditional explanation of the Tully-Fisher relation supposes that
light is proportional to mass: $L \sim M$.  This works if, among other
conditions, there is a universal baryon fraction.  Any distribution in
the baryon fraction should be reflected in the intrinsic scatter of
the relation.  The small observed scatter directly implies a narrow
$f_b$ distribution.  This restriction applies to all galaxies which
fall on the Tully-Fisher relation:  both the central galaxies of the
satellite studies and low surface brightness galaxies, and presumably the
progenitors of tidal tail systems as well.

We can quantify this limit by supposing that the Tully-Fisher relation
\begin{equation}
L \sim V^x
\end{equation}
arises from an underlying relation of the form
\begin{equation}
\mass \sim V^y.
\end{equation}
Presumably, $3 < x \approx y < 4$.
Note that some rather gross assumptions go into equation 2,
and fine-tuning problems involving the surface brightness or scale length
are unavoidable (Zwaan \etal 1995; McGaugh \& de Blok 1998).

Luminosity can be directly related to mass by
\begin{equation}
\mass = \frac{\ML* L}{f_* f_b}
\end{equation}
where \ML*\ is the mass-to-light ratio of the stars, $f_*$ is the fraction of
baryonic mass in the form of stars, and $f_b$ is the baryon fraction.
It follows that
\begin{equation}
\frac{\ML*}{f_* f_b} \sim V^{y-x}.
\end{equation}
By construction, the quantity $|y-x|$ must be small, $|y-x| < 1$.

For the moment, let us assume that \ML*\ and $f_*$ are finite but small
contributors to the scatter in the Tully-Fisher relation.  Even if
the $f_b$ distribution dominates the scatter,
\begin{equation}
\frac{\delta f_b}{f_b} < |y-x| \frac{\delta V}{V}.
\end{equation}
The intrinsic scatter in the Tully-Fisher relation tightly constrains
the allowed range of $f_b$.  For a generous assumptions of $|y-x| = 1$
and an intrinsic scatter of 0.5 mag.,
\begin{equation}
\frac{\delta f_b}{f_b} < 0.12.
\end{equation}
This is very small.  Moreover, we have ignored the scatter in \ML*\ and $f_*$.
From the perspective of stellar populations, one expects some scatter in \ML*.
Variation in $f_*$ is directly observed (McGaugh \& de Blok 1997).
This further tightens the constraint on the baryon fraction distribution.
This constraint specifically applies to the objects discussed above
where a factor of 2 or more variation was inferred, excepting only clusters of
galaxies.

{\bf This is a serious problem with no clear solution.}

\section{The Slope of the Tully-Fisher Relation}

We can also place limits on the amount by which the slope of
the observed Tully-Fisher Relation is allowed to vary from the underlying
mass-velocity relation.  Let us assume simply that the various components
are a function of luminosity:  $\ML* \sim L^a$, $f_* \sim L^b$, and
$f_b \sim L^c$.  By equation 4, these slopes are related by
\begin{equation}
a-b-c = y-x.
\end{equation}
A reasonable limit is $|a-b-c| < 1$, and probably rather less.

Given that brighter galaxies tend to be redder, the stellar mass-to-light
ratio is probably a weakly increasing function of $L$: $a > 0$, with its
precise value depending on bandpass.
The stellar mass fraction $f_*$ is observed to increase with luminosity roughly
as $b \sim 0.2$ (McGaugh \& de Blok 1997).  The two tend to offset one
another, so they should cause only a mild deviation of the observed slope
from the intrinsic one as long as $a \sim b$.  The usual assumption
$L \sim \mass$ thus seems well justified as longs as
the baryon fraction is a universal constant.  If this is not the case,
any systematic variation of $f_b$ with luminosity feeds
directly into the slope.  Since the observed slope is near to a reasonable
intrinsic value (Tully \& Verheijen 1997), this also argues against
substantial variation in $f_b$.


\end{document}